\documentclass[%
 reprint,
 amsmath,amssymb,
 aps,
]{revtex4-2}

\usepackage{graphicx}
\usepackage{dcolumn}
\usepackage{bm}
\usepackage{hyperref}
\usepackage{float}
\usepackage{braket}

\begin{document}

\preprint{APS/123-QED}

\title{Modulation Transfer Spectroscopy of the $\textrm{D}_1$ Transition of Potassium:\\ Theory and Experiment.}

\author{A. D. Innes$^1$}%
 \author{P. Majumder$^1$}%
 \author{H. R. Noh $^2$}%
    \email{hrnoh@chonnam.ac.kr}
 \author{S. L. Cornish$^1$}%
   \email{s.l.cornish@durham.ac.uk}
\address{
\mbox{$^{1}$Department of Physics, Durham University, South Road, Durham, DH1 3LE, United Kingdom}
\mbox{$^{2}$Department of Physics, Chonnam National University,} \mbox{Gwangju,61186 , Korea.}
}%

\date{\today}

\begin{abstract}
\noindent We report on a study of modulation transfer spectroscopy of the $4\textrm{S}_{1/2}\rightarrow 4\textrm{P}_{1/2}$ (D\textsubscript{1}) transition of naturally abundant potassium in a room-temperature vapour cell. This transition is critical for laser cooling and optical pumping of potassium and our study is therefore motivated by the need for robust laser frequency stabilisation. Despite the absence of a closed transition, the small ground-state hyperfine splitting in potassium results in strong crossover features in the D\textsubscript{1} modulation transfer spectrum. To emphasise this we compare the D\textsubscript{1} and D\textsubscript{2} spectra of potassium with those of rubidium. Further, we compare our experimental results with a detailed theoretical simulation, examining different pump-probe polarization configurations to identify the optimal signals for laser frequency stabilisation. We find good agreement between the experiment and the theory, especially for the $\textrm{lin} \parallel \textrm{lin}$ polarization configuration.

\end{abstract}

\maketitle


\section{\label{sec:1}Introduction}

Experiments employing laser cooling require lasers that are frequency stabilised to better than the linewidth of the transition used for cooling. For alkali-metal atoms this corresponds to a frequency stability of $\lesssim$ 1 MHz. One method of achieving this is with a spectroscopy technique to obtain an error signal from an atomic transition. The error signal is then used as feedback to correct any frequency deviations of the laser. Some of the most common methods are dithering the current to extract the derivative of the saturated absorption signal \cite{Haroche1972,Preston1996,Rovera1994}, frequency modulation spectroscopy \cite{Bjorklund1980,Mandon2007}, dichroic atomic vapour laser locking (DAVLL) \cite{Corwin1998,Millett2006,McCarron2007,Harris2008}, far off resonance locking with the Faraday effect \cite{Marchant2011,Quan2016}, polarization spectroscopy \cite{Wieman1976,Torii2012,Yoshikawa2003} and modulation transfer spectroscopy (MTS) \cite{Shirley1982,McCarron2008}. 

MTS is a pump-probe spectroscopy technique that works by modulating the pump beam with an electro-optical modulator (EOM) to produce sidebands and carrier components. When the pump beam is overlapped with the probe beam in an atomic vapour cell, and near resonance with an atomic transition, a four-wave mixing process transfers the sidebands from the pump to the probe. The beating between the sidebands and the probe can then be detected by a fast photodiode. Demodulation of the photodiode signal leads to the MTS signal. MTS has two key advantages. Firstly, it generates a dispersive signal on a zero background; the zero crossing of which occurs when the laser is exactly on resonance with the associated transition. Secondly, the MTS signal is dominated by cycling transitions. This can be useful in cases where the hyperfine structure is too narrow to resolve with other spectroscopic techniques such as the $\textrm{S}_{1/2} \rightarrow \textrm{P}_{3/2}$ transition in bosonic potassium \cite{Mudarikwa2012}. This makes it an effective tool for laser frequency stabilisation.

A considerable body of theoretical and experimental work has been published on MTS, for example see \cite{Shirley1982,McCarron2008,Noh2011,Li2011,Preuschoff2018,Ito2000,Cheng2014,Sun2016,Long2018,Mudarikwa2012,Wu2018,Negnevitsky2013}. There has already been a comprehensive study on the D\textsubscript{2} lines of potassium \cite{Mudarikwa2012}. However, to our knowledge, very little work has been published on the $\textrm{D}_1$ lines of potassium either in theory or experiment; an MTS spectrum has been reported but in a wider study of simultaneously locking multiple lasers to a single cell \cite{Mihm2018}.

The small hyperfine splitting in the $4\textrm{P}_{3/2}$ state of bosonic potassium makes it difficult to resolve the individual hyperfine components of the D\textsubscript{2} transition spectroscopically and has a detrimental impact on the efficiency of laser cooling \cite{Landini2011}. In contrast, the hyperfine components of the D\textsubscript{1} transition are resolvable due to the simpler structure and slightly larger splitting. This transition has found important applications for gray molasses cooling \cite{Chen2016,Salomon2013} and degenerate Raman sideband cooling \cite{Grobner2017} and is useful for efficiently spin-polarizing a sample of potassium atoms.

The bosonic isotopes of potassium also have the interesting feature that the ground-state hyperfine splitting is smaller than the Doppler width of the D\textsubscript{1} and D\textsubscript{2} transitions in a room-temperature vapour. This leads to the existence of ground-state crossover resonances in pump-probe spectroscopy schemes. These crossovers are something that are notably absent in the D line spectra of rubidium and caesium \cite{McCarron2007,Wu2018A}. Similar ground-state crossover resonances have been previously observed in the D\textsubscript{2} transition of lithium \cite{Sun2016}. The presence of ground-state crossover features in potassium motivate a comparison between the MTS spectrum of the D\textsubscript{1} and D\textsubscript{2} lines, as well as a comparison with a species, such as rubidium, whose ground-state hyperfine splitting is greater than the Doppler width. 

In this work we present a detailed study of MTS of the D\textsubscript{1} transition of potassium, comparing our results with MTS of the D\textsubscript{2} transition. To elucidate the role of the ground-state crossover, we contrast our results with spectra obtained for rubidium. We also present spectra for different pump-probe polarization configurations, showing that the configuration where the beams have linear and perpendicular polarization offers the strongest signal for locking. Throughout we compare our experimental results with the predictions from a theoretical model based upon the solutions to the time dependent optical Bloch equations without the use of any phenomenological constants. We find the calculations predict a large dispersive signal for the crossover features, as observed experimentally.

The layout of the paper is as follows. In Sec. \ref{sec:2} we present the theoretical model used to predict the MTS spectrum. In Sec. \ref{sec:3} we outline the details of our experiment. In Sec. \ref{sec:4} we present our results and compare them against the predicted theoretical curves. In Sec. \ref{sec:5} we summarise our work and give an outlook to possible future extensions of our study.

\section{\label{sec:2}Theory}

\begin{figure}[t!]
    \centering
    \includegraphics[width = 0.47\textwidth]{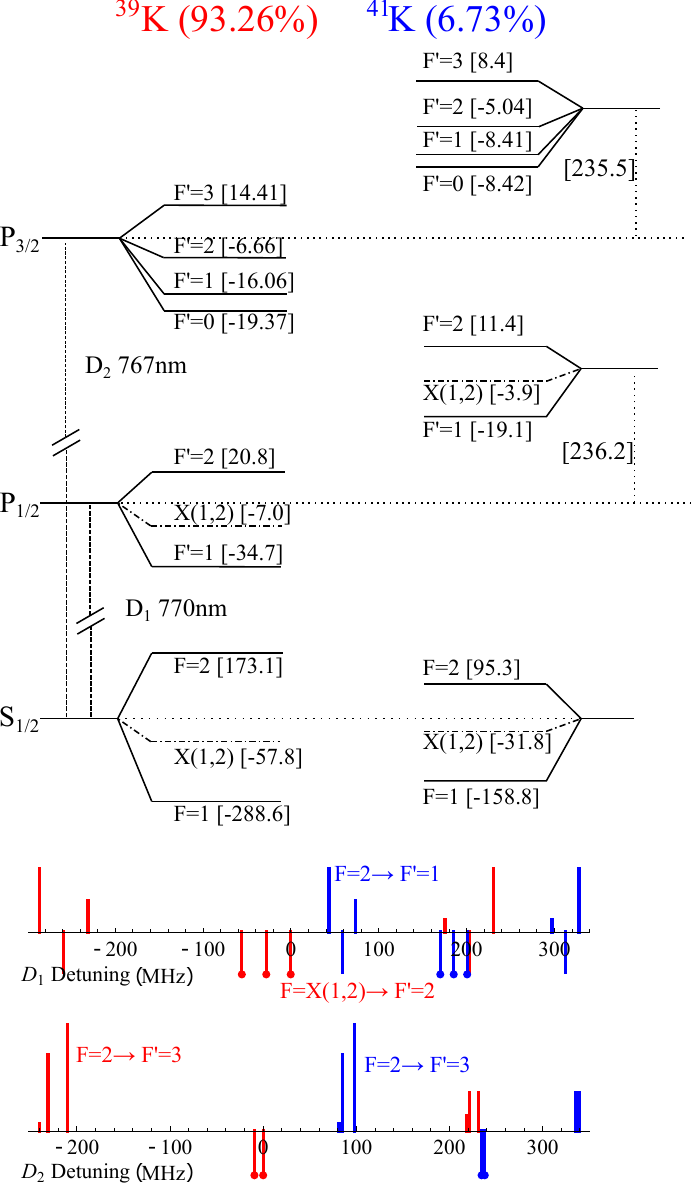}
    \caption{The hyperfine structure of potassium relevant to the D\textsubscript{1} and D\textsubscript{2} transitions, together with the associated line spectra highlighting the relative transition frequencies. Levels are labelled by the quantum number $F$ for the total angular momentum of the state and hyperfine shifts are given in MHz using the values from \cite{Tiecke}. The horizontal dot-dashed lines show the positions of crossovers for the S\textsubscript{1/2} and P\textsubscript{1/2} manifolds. In the line spectra, lines above the axis correspond to normal transitions and have heights reflecting the relative oscillator strengths. Lines below the axis correspond to crossover transitions. Ground-state crossovers are indicated by circles at the  bottom of the line. For clarity, excited-state crossover transitions have been omitted for the D\textsubscript{2} transition. For both spectra, zero detuning corresponds to the $X(1,2) \rightarrow F'=2$ transition in \textsuperscript{39}K. Note, in the experiment the spectra will be weighted by the natural abundances of the isotopes shown in the figure.}
    \label{Hyperfine}
\end{figure}

In this section, we outline the theory for predicting the MTS signal. Figure \ref{Hyperfine} shows the hyperfine structure for the 4S\textsubscript{1/2}, 4P\textsubscript{1/2} and 4P\textsubscript{3/2} energy levels for the bosonic isotopes. We illustrate the method of calculation for the D\textsubscript{1} transition in \textsuperscript{39}K, noting that the methodology is directly applicable to \textsuperscript{41}K as both isotopes have the same nuclear spin. The final spectra are obtained by summing the results for \textsuperscript{39}K and \textsuperscript{41}K weighted by their natural abundances. We consider four polarization configurations of the counter-propagating pump and probe laser beams: (i) $\textrm{lin} \| \textrm{lin}$, (ii) $\textrm{lin} \bot \textrm{lin}$, (iii) $\sigma^+\sigma^+$, and (iv) $\sigma^+\sigma^-$. Here, for simplicity, the circularly polarized cases in (iii) and (iv) are labeled by the transitions the beams drive. In the case of linear polarization configurations, the quantization axis was chosen as the direction of the electric field of the pump beam, whereas the propagation direction of the pump beam was chosen as the quantization axis in the case of circular polarization configurations. In the experiment, we reinforce the quantization axis with a weak magnetic field parallel to the direction of beam propagation. The polarization vectors of the pump and probe beams are expanded as $a_+ \hat{\epsilon}_+ +a_0 \hat{\epsilon}_0 +a_- \hat{\epsilon}_-$ and $c_+ \hat{\epsilon}_+ +c_0 \hat{\epsilon}_0 +c_- \hat{\epsilon}_-$, respectively, in the spherical bases. Then, the coefficients for the $\pi$- and $\sigma^+$ polarized pump beam are $(a_+,a_0,a_-) = (0,1,0)$ for polarization configurations (i) and (ii) and $(1,0,0)$ for configurations (iii) and (iv). The corresponding coefficients $(c_+,c_0,c_-)$ for the probe beam are (i) $(0,1,0)$, (ii) $(- 1/\sqrt{2},0,1/\sqrt{2})$, (iii) $(1,0,0)$, and (iv) $(0,0,1)$.

To find the internal dynamics of the atoms, we solve the following density matrix equation:
\begin{eqnarray}\label{density}
\dot \rho = -(i/\hbar) \left[ H_0 +V ,\rho \right] +{\dot \rho}_{\rm relax},
\end{eqnarray}
where $\rho$ is the density operator, and $H_0$ and $V$ are the atomic and interaction Hamiltonians, respectively. In Eq. (\ref{density}), ${\dot \rho}_{\rm relax}$ represents the term related to relaxations such as spontaneous emission and transit-time decay \cite{Choi2016}. The atomic and interaction Hamiltonians are given by
\begin{eqnarray}\label{H0}
H_0 &=& -\sum_{m'=-2}^{2} \hbar  \delta_1  \left| F'=2,m' \right> \left< F'=2,m' \right|  \nonumber \\
&&-\sum_{m'=-1}^{1} \hbar \left( \delta_1 +\Delta_e  \right) \left| F'=1,m' \right> \left< F'=1,m' \right|  \nonumber \\
&&- \sum_{m=-1}^{1}\hbar \Delta_g \left| F=1,m \right> \left< F=1,m \right|   , \\
\label{V}
V &=& \sum_{F=1}^2 \sum_{F'=1}^2 \sum_{m =-F}^{F} \sum_{q=\pm,0} \frac{\hbar}{2} \left[ c_q \Omega_p e^{-i \delta_p t} \right.\nonumber \\
&& \left.  +a_q ( \Omega_c +\Omega_s e^{-i \Omega t} -\Omega_s e^{i \Omega t} ) \right] \nonumber \\ && \times C_{F,m}^{F',m'+q} \left| F',m' +q \right> \left< F,m \right| +{\rm h.c.} ,
\end{eqnarray}
respectively, where  h.c. denotes the Hermitian conjugate. 

In Eqs. (\ref{H0}) and (\ref{V}), $\delta_1 ( \equiv \delta+k v)$ is the detuning of the carrier component of the pump beam. $\delta_p$($=-2kv$) is the detuning of the probe beam relative to the carrier component felt by an atom moving at velocity $v$ where $\delta$ is the detuning of the laser frequency with respect to $F =2 \rightarrow F'=2$ transition of \textsuperscript{39}K and $k$ is the wave vector. $F$ ($F'$) and $\Delta_g$ ($\Delta_e$) denote the hyperfine quantum number and splitting of the ground (excited) state, respectively. $m$ and $m'$ are the projection of the total angular momentum onto the quantisation axis for the ground and excited states, respectively. In Eq. (\ref{H0}), the external magnetic field was not taken into consideration. In Eq. (\ref{V}), $\Omega_c$ and $\Omega_s$ represent the Rabi frequencies of the carrier and sideband components of the pump beam, respectively; $\Omega_p$ is the Rabi frequency of the probe beam. $\Omega$ is the modulation frequency for the sidebands, and $C_{F,m}^{F',m'}$ is the normalized transition strength between the states $\left|F,m\right>$ and $\left|F',m'\right>$ \cite{Noh2011}. In Eq. (\ref{V}), $c_q$ ($a_q$) with $q=\pm, 0$ are the coefficients of the electric field vector of the probe (pump) beam in the spherical bases, as mentioned above.

To solve Eq. (\ref{density}) using the explicit expression of Hamiltonians given in Eqs. (\ref{H0}) and (\ref{V}) in the steady-state regime, the density matrix elements must be expanded as various Fourier components oscillating in time with various oscillation frequencies. The detailed description of finding oscillation frequencies at general polarization configurations were reported in Refs. \cite{Noh2011,Lee2021}. Thus, when the three-photon interactions are considered, the optical coherences have 20 oscillation frequencies and Zeeman coherences and populations have 11 oscillation frequencies in the cases of $\textrm{lin} \| \textrm{lin}$ and $\sigma^+\sigma^+$ configurations \cite{Lee2021}. We may use these frequencies in the cases of $\textrm{lin} \bot \textrm{lin}$ and $\sigma^+\sigma^-$ configurations as well. However, we can select further non-vanishing components of the density matrix elements. For example, when the polarization configuration is $\sigma^+\sigma^-$, the oscillation frequencies for the optical coherences are given by $ \left\{-\delta_p, -\delta_p \pm \Omega, -\delta_p \pm 2 \Omega \right\} $ and $\left\{0, \pm \Omega , \pm 2 \Omega, \pm 3 \Omega  \right\}$ for $\Delta m (\equiv m'-m) =-1$ and $+1$, respectively. Those frequencies for $\Delta m = -3$ and $+3$ are given by $\left\{ -2 \delta_p , -2 \delta_p \pm \Omega  \right\}$ and $\left\{\delta_p , \delta_p \pm \Omega ,  \delta_p \pm 2 \Omega  \right\}$, respectively. In the $\sigma^+\sigma^-$ configuration, the optical coherences not satisfying $\Delta m = \pm 1$, and $\pm 3$ vanish. The oscillation frequencies for the populations are  $\left\{ 0, \pm \Omega, \pm 2 \Omega  \right\}$. We will not present the description of Zeeman coherences for the $\sigma^+\sigma^-$ configuration nor the optical and Zeeman coherences for the $\textrm{lin} \bot \textrm{lin}$ configuration.

Using the expanded density matrix elements, a set of coupled differential equations for Fourier components of the density matrix elements is obtained from Eq.~(\ref{density}), which is then solved in a steady-state regime as functions of $v$ and $\delta$. The MTS signal can be obtained from the relevant optical coherences with oscillation frequencies of $-\delta_p \pm \Omega$, whose real and imaginary parts are defined as $r_{F,m}^{F',m',(\pm)}$ and $s_{F,m}^{F',m',(\pm)}$, respectively. Then, the in-phase ($I_n$) and quadrature ($Q_n$) components of the MTS signals are given by
\begin{eqnarray}\label{I}
&&I_{n} =  \sum_{F=1}^2 \sum_{F'=1}^2 \sum_{m =-F}^{F} \sum_{q= \pm, 0} 
c_q C_{F,m}^{F',m+q}  \nonumber \\
&&\quad  \times  \int_{-\infty}^{\infty} {\rm d}v f_{\rm D}(v)  \left(s_{F,m}^{F',m+q,(-)} +s_{F,m}^{F',m+q,(+)} \right) , \\
&&Q_{n} =  \sum_{F=1}^2 \sum_{F'=1}^2 \sum_{m =-F}^{F} \sum_{q= \pm, 0} 
c_q C_{F,m}^{F',m+q}  \nonumber \\
&& \quad \times \int_{-\infty}^{\infty} {\rm d}v f_{\rm D}(v)  \left(-r_{F,m}^{F',m+q,(-)} +r_{F,m}^{F',m+q,(+)} \right) \label{Q}, 
\end{eqnarray}
where $n=39$ and 41 represent $^{39}$K and $^{41}$K, respectively, and atomic parameters for specific isotopes are used in the calculation. In Eqs. (\ref{I}) and (\ref{Q}), $f_{\rm D}(v) [=1/(\sqrt{\pi}u)  e^{-(v/u)^2}]$ represents the Maxwell--Boltzmann velocity distribution function and $u$ is the most probable speed of the atom in the cell. The final MTS signals including both $^{39}$K and $^{41}$K are given by
\begin{eqnarray}
I(\delta) &=& \frac{N_{39}}{N_{39}+N_{41}} I_{39} (\delta) +\frac{N_{41}}{N_{39}+N_{41}} I_{41} (\delta -\Delta), \\
Q(\delta) &=& \frac{N_{39}}{N_{39}+N_{41}} Q_{39} (\delta) +\frac{N_{41}}{N_{39}+N_{41}} Q_{41} (\delta -\Delta) ,
\end{eqnarray}
where $N_{39(41)}$ is the atomic density of $^{39}$K($^{41}$K) in the cell considering the natural abundances of the isotopes and $\Delta =2 \pi \times$ 305 MHz is the isotope shift of the $F=2 \rightarrow F'=2$ transition between $^{39}$K and $^{41}$K.

\section{\label{sec:3}Experiment}

\begin{figure*}[t!]
    \centering
    \includegraphics{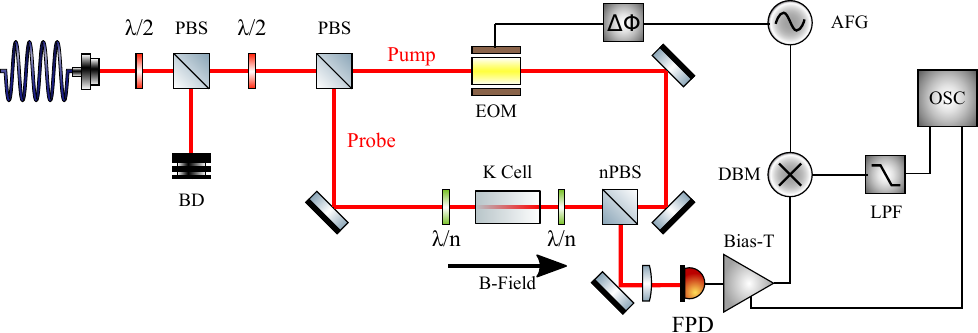}
    \caption{Experimental setup for modulation transfer spectroscopy (MTS). Laser light is output from a fiber and the red solid line indicates the subsequent beam path. Two polarizing beam splitters (PBS) and waveplates are used to set the total power and ratio between the pump and probe beams. The waveplates labelled $\lambda/n$ are exchanged depending on the polarization configuration: $n=2$ for linear polarization configurations or $n=4$ for circular polarization configurations. The transmitted probe beam is detected on a fast photodiode (FPD). The resulting MTS signal is extracted using a double-balanced mixer (DBM) and low-pass filter (LPF). nPBS: 50:50 non-polarizing beam splitter. BD: Beam dump. EOM: Electro-Optical modulator. AFG: Arbitrary function generator. OSC: Oscilloscope. }
    \label{setup}
\end{figure*}

The experimental setup is shown in Fig. \ref{setup}. The laser source is an external cavity diode laser (Toptica DL Pro). We couple the laser light through a single mode polarization maintaining fiber such that the subsequent output beam profile is Gaussian.  A pair of polarizing beam splitters (PBS) in combination with a pair of $\lambda/2$ waveplates are used to control the total power of laser light delivered to the spectroscopy setup and the ratio of power between the pump and probe beams.  The probe light is passed directly to the 2\,cm long potassium vapour cell. The cell is housed inside a brass block with a pair of heating elements attached. The cell temperature is raised to 99(2)\textdegree C where we expect a vapour pressure of $3.5 \times 10^{-5}$ mbar. The probe and pump beams are collimated such that their $1/e^2$ diameters at the centre of the cell are $2.06(2)\,\textrm{mm}$ and $1.96(5)\,\textrm{mm}$, respectively.

The pump light is passed through a homebuilt EOM. The EOM uses a LiTaO\textsubscript{3} crystal electrically contacted to a pair of brass capacitor plates. The addition of an inductor creates a simple LCR circuit that resonantly enhances the voltage across the crystal. The resonance frequency of the EOM is at 6.054(5)~MHz. We always drive the EOM on resonance at its maximum voltage, producing sidebands each with intensities equal to $15(1)\% $ of the total pump intensity. More details of the EOM can be found in previous work~\cite{McCarron2007}.

We investigate different combinations of laser polarization. To maintain a well defined quantisation axis we apply a weak magnetic field with a set of rectangular coils which are concentric with the cell. The coils provide a 1.5 G magnetic field at the centre of the cell in a direction along the beam propagation axis. By switching around the waveplates indicated in Fig. \ref{setup} we are able to study four different polarization configurations of the laser light through the cell: circular polarization where the pump and probe drive opposite transitions ($\sigma^+\sigma^-$), circular polarization where the pump and probe drive the same transitions ($\sigma^+\sigma^+$), linear polarization where the pump and probe are perpendicular ($\textrm{lin}\bot\textrm{lin}$) and linear polarization where the pump and probe are parallel ($ \textrm{lin} \parallel \textrm{lin} $). 

A simple set of electronics apply a phase lock-in technique to demodulate the MTS signal. The homebuilt fast photodiode detects the beat signal between the modulated probe carrier and the sidebands. This signal is passed through a bias-tee [Mini-Circuits model:
ZFBT-4R2GW] which filters out the DC component of the signal corresponding to the standard saturated absorption profile and sends it to a secondary channel of the oscilloscope. The modulated component is passed to the double-balanced mixer (DBM) [Mini-Circuits model: ZAD-1H+] which multiplies the signal with the reference signal provided by the arbitrary function generator (AFG) [Tekronix model: AFG 3102]. The AFG has two outputs. One output is used to drive the EOM and the second output is used as a local oscillator which acts as our reference for the demodulation. The relative phase of the two outputs can be set using the AFG, allowing us to fully characterise the phase parameter. The demodulated signal can then be sent to the oscilloscope for data acquisition or to feedback circuitry to stabilize the laser frequency.

\begin{figure*}[t!]
    \centering
    \includegraphics[width = \textwidth]{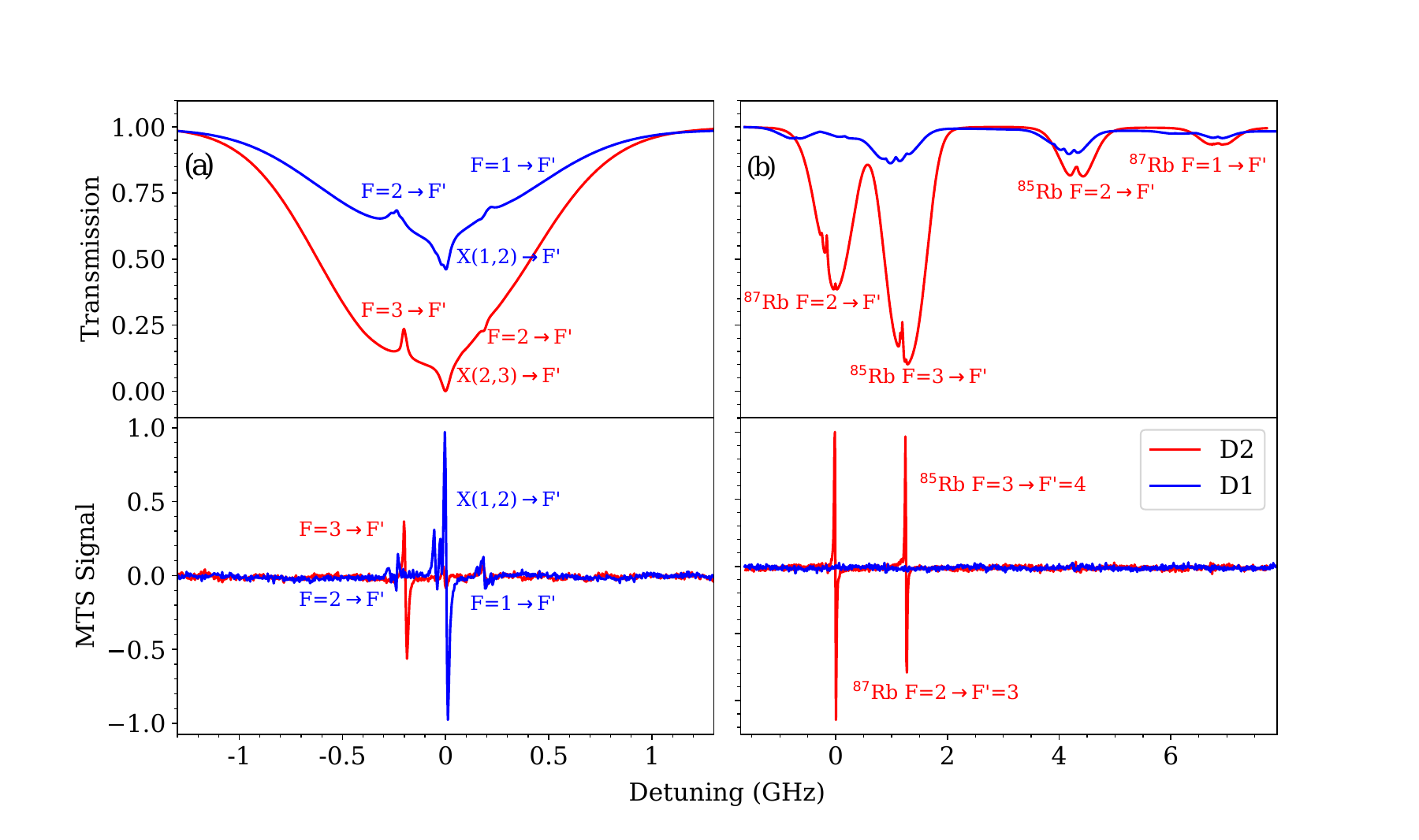}
    \caption{(a) Absorption signal as filtered by bias-T (top) and MTS signal (bottom) for potassium in the $\sigma^+\sigma^+$ configuration. The vapour cell was 2~cm long and heated to $99~^\circ \textrm{C}$. Data were taken with probe and pump beam intensities of $78(2)~\textrm{mW cm}^{-2}$ and $86(5)~\textrm{mW cm}^{-2}$ respectively. (b) Signal for rubidium in the $\sigma^+\sigma^+$ configuration. The vapour cell was 4~cm long and at $20~^\circ \textrm{C}$. Data were taken with probe and pump beam intensities of $90(2)~\textrm{mW cm}^{-2}$ and $53(2)~\textrm{mW cm}^{-2}$ respectively. The red lines show the D\textsubscript{2} transition and the blue lines show the D\textsubscript{1} transition. For potassium, zero detuning corresponds to the $X(1,2) \rightarrow F'=2$ transition in \textsuperscript{39}K for the D\textsubscript{1} trace and the $X(2,3) \rightarrow F'=3$ transition for the D\textsubscript{2} trace. For rubidium, zero detuning corresponds to the $F=2 \rightarrow F'=2$ in \textsuperscript{87}Rb for the D\textsubscript{1} trace and the $F=2 \rightarrow F'=3$ transition for the D\textsubscript{2} trace. The vertical axes for both (a) and (b) share the same ticks.}
    \label{fig:D2}
\end{figure*}

\section{\label{sec:4}Results}

\subsection{Comparison with D\textsubscript{2} and rubidium MTS}
To better illustrate the role of the crossovers and cycling transitions in MTS we have recorded an MTS trace for the D\textsubscript{2} and D\textsubscript{1} transitions of both potassium and rubidium for a brief comparative study. \textsuperscript{87}Rb is suitable for a comparison against potassium since they both have the same hyperfine quantum numbers.

Figure~\ref{fig:D2}(a) shows the absorption spectroscopy and MTS signals for the D\textsubscript{1} and D\textsubscript{2} transitions in potassium. The vertical axis for the MTS signals are normalised with respect to the \textsuperscript{39}K D\textsubscript{1} $X(1,2) \rightarrow F=2$ feature. These data were taken using the $\sigma^+\sigma^+$ configuration with a probe beam intensity of $78(2)~\textrm{mW cm}^{-2}$ and a pump beam intensity of $86(5)~\textrm{mW cm}^{-2}$. All intensities quoted for the experiment are the peak intensity where the beam is assumed to be Gaussian. We note from the absorption profile on Fig.~\ref{fig:D2}(a) that all of the hyperfine transitions lie within the same Doppler profile. The D\textsubscript{2} transition shows the standard MTS signal, with a single strong feature resulting from the $F=2 \rightarrow F'=3$ closed transition on a flat zero background. This feature is ideal for laser frequency stabilization~\cite{Mudarikwa2012}. For the D\textsubscript{1} transition we observe several strong features in the MTS signal, despite the absence of a closed transition. The signal is dominated by the ground-state crossover transition, $X(1,2) \rightarrow F'=2$, providing a feature suitable for laser frequency stabilization. We note that the $X(1,2)$ crossover features involve both ground-state energy levels such that the pump beam and probe beam can simultaneously interact with atoms in both states preventing hyperfine optical pumping, effectively `closing' the transition. Such strong ground-state crossovers have also been reported in the D\textsubscript{2} MTS of lithium \cite{Sun2016}. In contrast, the crossover signals for the D\textsubscript{2} transition are weak, as reported previously \cite{Mudarikwa2012}. For the $\sigma^+\sigma^+$ configuration the existence of a cycling transition may be detrimental; optical pumping on the cycling transition will populate the $F=2, m_f=2$ ground state, but on the crossover at least one of the pump or probe beams will be interacting with the depopulated $F=1$ state. This will restrict transfer of the sidebands from the pump to the probe, weakening the signal. The strong resonances observed for MTS of the D\textsubscript{1} transition in the absence of a cycling transition merit further study.

Figure~\ref{fig:D2}(b) shows the absorption spectroscopy and MTS signals for the D\textsubscript{1} and D\textsubscript{2} transitions in rubidium. These data were taken taken using the $\sigma^+\sigma^+$ configuration with a probe probe beam intensity and pump beam intensity of $90(2)~\textrm{mW cm}^{-2}$ and $53(2)~\textrm{mW cm}^{-2}$, respectively. Again the MTS signal for the D2 transition is dominated by the cycling transitions. However, in contrast to K, the MTS signal on the D\textsubscript{1} transitions for Rb does not  show any measurable features. This can be attributed to the lack of ground-state crossover transitions; in Rb the ground-state hyperfine splitting (greater than 3 GHz for both isotopes) is significantly greater than the Doppler broadening (on the order 500~MHz in a room temperature cell), as evident in the absorption spectrum.

\subsection{MTS signal optimisation and comparison with theory}

\begin{figure*}[t!]
    \centering
    \includegraphics[width = 0.9\textwidth]{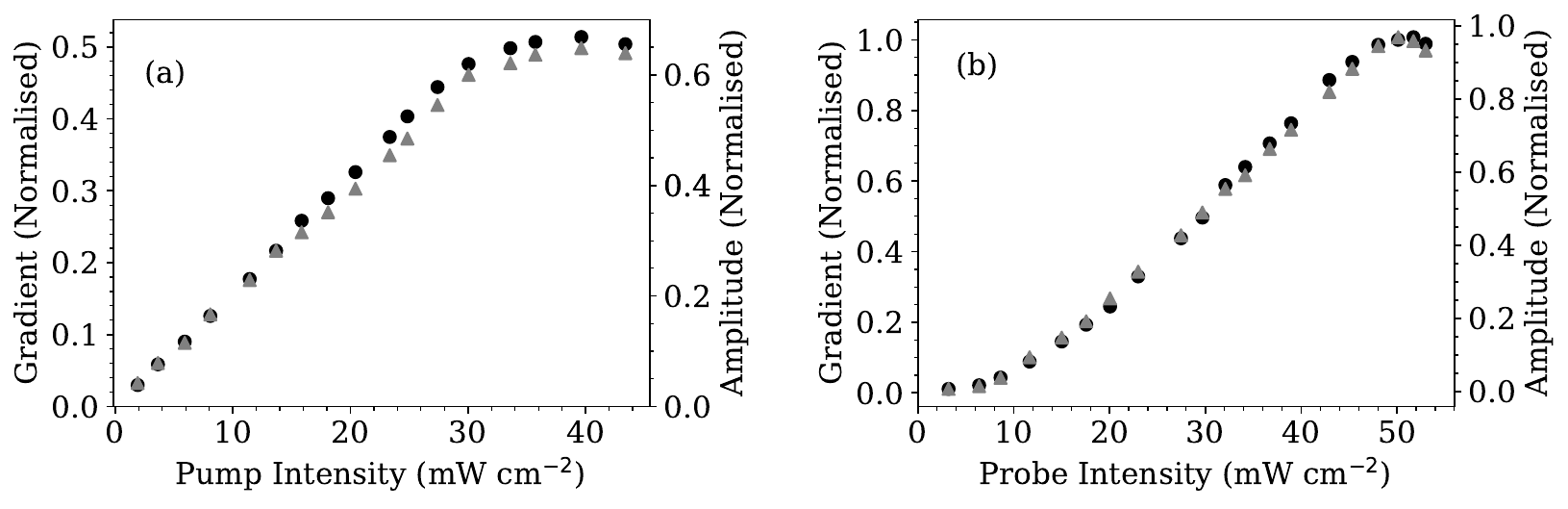}
    \caption{Gradient (black circles) and amplitude (grey triangles) of the $X(1, 2) \rightarrow F'= 2$ \textsuperscript{39}K transition as a function of the beam intensity in the $\textrm{lin}\parallel\textrm{lin}$ configuration. Results are shown for (a) the pump beam optimised at a constant probe of $47.4(8)~\textrm{mW cm}^{-2}$ and (b) the probe beam optimised with a constant pump of $32.8(2)~\textrm{mW cm}^{-2}$.}
    \label{fig:Opt}
\end{figure*}

\begin{figure*}[t!]
    \centering
    \includegraphics[width = 1\textwidth]{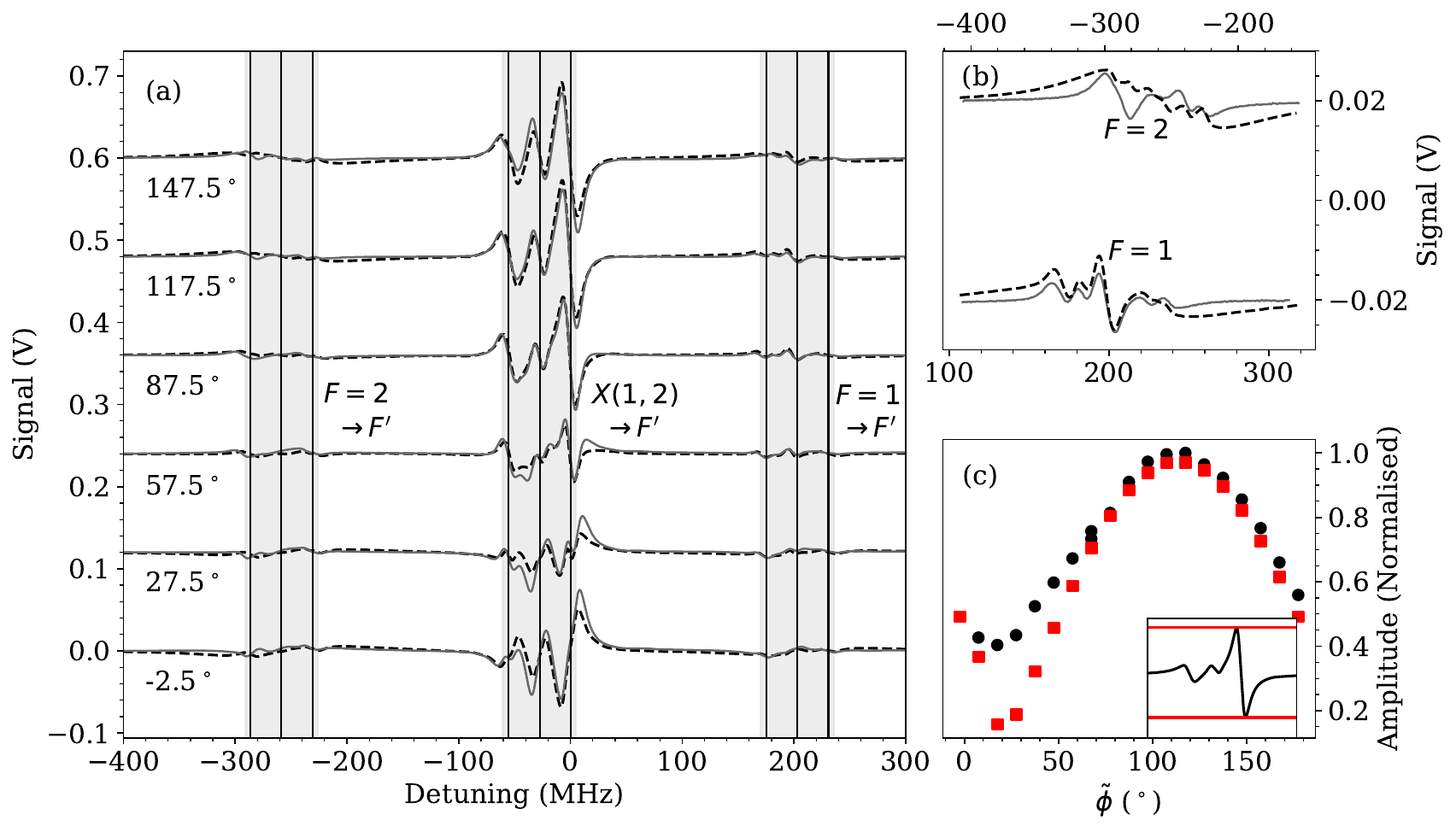}
    \caption{(a) MTS signals on the potassium D$_1$ transition for the $\textrm{lin} \parallel \textrm{lin}$ configuration for different relative phases at the double-balanced mixer. The data were taken with intensities for the pump carrier, the pump sidebands and the probe of 25(2)$\textrm{mW cm}^{-2}$, 5.5(4)$\textrm{mW cm}^{-2}$ and 59.5(1.1)$\textrm{mW cm}^{-2}$, respectively. The grey solid trace is the experimental data and the black dashed trace is the fitted simulation. An arbitrary vertical offset is added between traces to separate the signals recorded at different phases. Highlighted regions indicate areas where we see transitions and have their ground-state shown next to them. Vertical lines show the transition locations for \textsuperscript{39}K. Numbers on the left of each phase show the phase, $\tilde{\phi}$. (b)  MTS signal zoomed into the $F=2\rightarrow F'$ (top) and $F=2 \rightarrow F'$ (bottom) transitions at $\tilde{\phi}=117.5~^\circ$. A vertical offset has been added to separate the two transitions. (c) The amplitude of the $X(1,2) \rightarrow F'=2$ feature as a function of the phase for experimental (black circles) and theoretical predictions (red squares).}
    \label{LPaLPhase}
\end{figure*}

\begin{figure*}[t!]
    \centering
    \includegraphics[width = 0.9\textwidth]{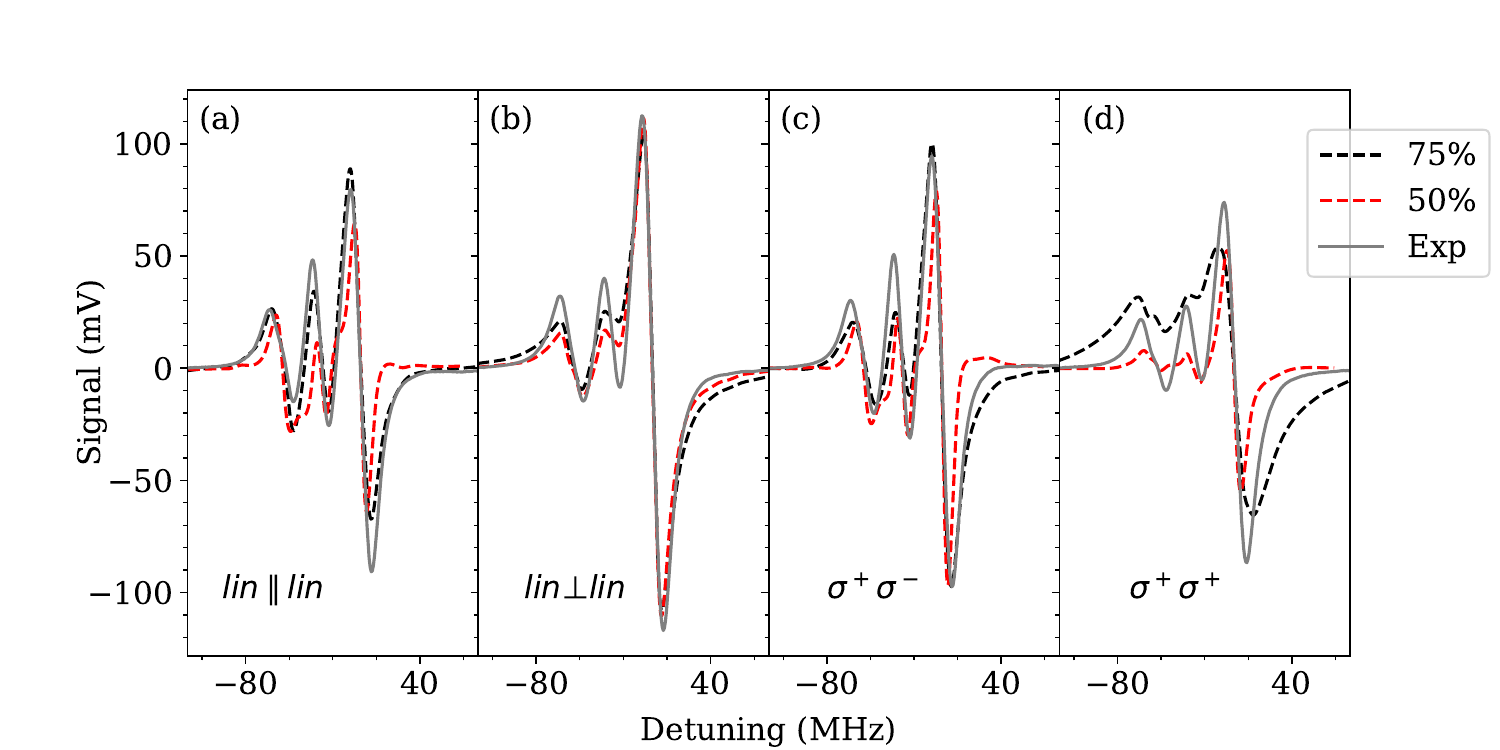}
    \caption{The MTS signal on the potassium D$_1$  $X(1,2) \rightarrow F'$ transitions for different polarization configurations. All the traces were recorded using the optimum parameters for the $\textrm{lin} \parallel \textrm{lin}$ configuration. The gray solid lines show the experimental data and the black lines indicate the best fit theory for simulations at 75\% intensities. We have also included, for comparison, the best fit theory data for a simulation at 50\% of the quoted intensities: the red dashed lines. The polarization configurations are (a) $\textrm{lin} \parallel \textrm{lin}$, (b) $\textrm{lin} \perp \textrm{lin}$, (c) $\sigma^+ \sigma^-$ and (d) $\sigma^+ \sigma ^+.$}
    \label{AllPol}
\end{figure*}
We use measurements of the strongest feature resulting from the $X(1,2) \rightarrow F'=2$ transition to optimise the parameters of the MTS setup for the D1 transition in potassium. Figure \ref{fig:Opt} shows the gradients and amplitudes of the signal for the $\textrm{lin}\parallel\textrm{lin}$ configuration as a function of the intensities of (a) the pump beam and (b) the probe beam, with the intensity of the other beam held constant. We optimise the intensities to obtain the steepest gradient. The pump intensity was optimised first with a constant probe intensity of 47.4(8)\,$\textrm{mW cm}^{-2}$; then the probe intensity was optimised with a constant pump intensity of 32.8(2)\,$\textrm{mW cm}^{-2}$ (chosen arbitrarily). The optimised intensities were found to be $36(3)~ \textrm{mW}~\textrm{cm}^{-2}$, and $59.5(1.1) ~\textrm{mW}~\textrm{cm}^{-2}$ for the pump and probe beams, respectively. Note the saturation intensity is $1.75~\textrm{mW}~\textrm{cm}^{-2}$. 

In Fig.~\ref{LPaLPhase} we investigate the dependence of the MTS signal for the $\textrm{lin} \parallel \textrm{lin}$ configuration as a function of the relative phase, $\tilde{\phi}$, at the double-balanced mixer. Traces of different phase were taken by introducing a phase delay using the AFG and were taken in $10^\circ$ steps. Theoretical traces were calculated as discussed in Sec. \ref{sec:2}. The simulation assumed a flat-top intensity distribution rather than the Gaussian profile used in the experiment. To partially compensate for this, intensities used in the simulation were set to 75\% of the quoted peak experimental intensities. The width of this top-hat function is equal to the $1/e^2$ diameter of the beam. Two contributions are calculated: an in-phase component $I(\delta)$ and a quadrature component $Q(\delta)$, where $\delta$ is the detuning. These are superimposed to obtain the signal, $S$, as a function of the relative phase $\phi$,
\begin{equation}
    S[\delta,\phi] = A(I(\delta)\cos\tilde{\phi} + Q(\delta)\sin\tilde{\phi}),
    \label{eqn:TotalSignal}
\end{equation}
\noindent where $A$ is an arbitrary amplitude factor to match the theory to the experimental measurements. Experimentally there are two contributions to $\tilde{\phi}$; the relative phase $\phi$ set by the AFG and an arbitrary phase offset $\phi_0$ arising from cable delays. Thus to compare the theory and experiment we use $\tilde{\phi} = \phi - \phi_0$, and adjust the value of $\phi_0$ to find the best agreement. In practice, traces for all measured phases of a given polarization are fitted simultaneously to Eq. (\ref{eqn:TotalSignal}) to extract the parameters $A$ and $\phi_0$.

Figure \ref{LPaLPhase}(a) shows example MTS signals for both the experiment (solid lines) and theoretical predictions (dashed lines) in phase steps of $30^{\circ}$. There are three distinct regions highlighted corresponding to features arising from the $F=1$ ground state (right), the $X(1,2)$ crossover (centre) and the $F=2$ ground state (left). It is also worth noting that the rightmost region also has a significant contribution from the \textsuperscript{41}K $X(1,2)$ crossovers as well, see Fig. \ref{Hyperfine}, but the other features are purely from \textsuperscript{39}K.
Figure \ref{LPaLPhase}(b) shows a zoom in to the $F=2 \rightarrow F'$ (top) and $F=1 \rightarrow F'$ (bottom) ground-state transitions, respectively. The peak-to-peak amplitude for the $X(1,2) \rightarrow F' =2$ feature as a function of the phase is shown in Fig. \ref{LPaLPhase}(c). We note that the optimal peak-to-peak amplitude is not obtained in either the quadrature or in-phase scenarios, but rather at a relative phase of $\tilde{\phi} \sim 112^\circ$.

We can see from Fig.~\ref{LPaLPhase} that there is generally good agreement between theory and experiment. For phases where the $X(1,2) \rightarrow F'$ features are maximised, the relative peak heights and signal widths agree particularly well. Further evidence is shown by (c) which compares the experimental amplitude against the predicted amplitude for the dominant feature over phases from $0^\circ$ to $180^\circ$. We certainly see in the region closest to the maximum that we get better agreement. For the weaker $F=1 \rightarrow F'$ and $F=2 \rightarrow F'$ transitions shown in (b) there is reasonable agreement, although there are differences in some cases.

\subsection{Different polarization configurations}

A simple reconfiguration of the waveplates in Fig. \ref{setup} allows us to look at the MTS signal for other polarization configurations. The results are shown in Fig.~\ref{AllPol}  for the signals arising from $X(1,2) \rightarrow F'$ transitions. The traces were taken with $\tilde{\phi} = 117.5~^\circ$, close to the optimum phase of $\sim 112~^\circ$ degrees from Fig.~\ref{LPaLPhase}(b). The laser parameters were the same as the optimum values found for the $\textrm{lin}\parallel\textrm{lin}$ configuration. In the $\textrm{lin} \perp \textrm{lin}$ case we observe a substantial increase in the amplitude of the $X(1,2) \rightarrow F'=2$ signal and a reduction in the size of the neighbouring features. There is very good agreement between the theory and experiment for all cases except for the $\sigma^+\sigma^+$ configuration. We note that in this case the theory predicts substantially broader features than are observed experimentally. The Rabi frequencies of laser beams in the experiment lie between 2--4$\times\Gamma$. In this region of laser intensities, our assumption of three-photon interactions may be insufficient, in particular for the specific polarization configuration, and we may need new calculations of higher order interactions in this case. This kind of improvement in the calculation is beyond the scope of the current paper.

\begin{table}[h!]
\vspace{0.1in}
\begin{tabular}{|l|l|l|}
\hline
Polarization         & $A$ (V)    & $\phi_0$ (\textdegree)  \\ \hline
$\textrm{lin} \parallel \textrm{lin}$      & $1787(7)$ & $110.7(4)$ \\ \hline
$\textrm{lin} \perp \textrm{lin}$ & $886(3)$  & $109.4(3)$ \\ \hline
$\sigma^+\sigma^-$   & $1576(5)$ & $108.3(3)$ \\ \hline
$\sigma^+\sigma^+$   & $1232(8)$ & $103.7(7)$ \\ \hline
\end{tabular}
\caption{Fitted parameters for $\textrm{D}_1$ experimental results against theory. Each of the polarization configurations are listed. $A$ and $\phi_0$ are as defined in Eq. (\ref{eqn:TotalSignal}).}
\label{tab:D1results}
\end{table}

Table \ref{tab:D1results} shows the fitted values of $A$ and $\phi_0$ for all polarization configurations. The values of $A$ and $\phi_0$ mean little on their own since they are arbitrary and exclusive to our specific setup, however in the context of theory we expect these values to be constant across different polarization configurations. We find good agreement between the different configurations for the phase offset, with all cases falling within a $7~^\circ$ range ($1.4^\circ$ if we exclude the $\sigma^+\sigma^+$ data). The amplitude conversion factor, $A$, does not agree between different configurations. We note that although the $\textrm{lin} \parallel \textrm{lin}$ and the $\sigma^+ \sigma^-$ results are are in reasonable agreement; the $\textrm{lin} \perp \textrm{lin}$ disagrees by a factor of 2 compared with $\textrm{lin} \parallel \textrm{lin}$. The theory predicts that the $X(1,2) \rightarrow F = 2$ transition to be at least a factor of two larger in the linear perpendicular case than in the $\textrm{lin} \parallel \textrm{lin}$ case. However, this is not observed in experiment.

It is of interest to investigate how the theoretical predictions depend upon the intensities used. As mentioned previously, the simulation assumes a uniform flat-top intensity distribution but the beam in the experiment is Gaussian. Figure \ref{AllPol} also includes the simulations for intensities at 50\% of the experimental peak intensities. Across all polarization configurations the broadening is substantially less. Although the relative heights of the peaks is somewhat mismatched we note an immediate improvement in the agreement between the theory and experiment for the $\sigma^+\sigma^+$. However, the width of the other features is underestimated by the theory in this case. This is a crucial point. If simulating the beam at the peak intensity quoted, then it would be a uniform beam but with a constant intensity at the peak of the Gaussian beam. This may result in an overestimation of power being delivered to the atoms. Likewise a uniform beam at 50\% of the peak intensity (the average beam intensity) may underestimate the power at the atoms. It is evident from the simulations shown in Fig. \ref{AllPol} that there is a strong intensity dependence from the theory which may assist in explaining some of discrepancies. We choose, albeit arbitrarily, 75\% of the peak beam intensity obtain theory between these two extremes. Regardless, the simulation still produces results in reasonable agreement with the experiment for the $\textrm{lin} \parallel \textrm{lin}$, $\textrm{lin} \bot \textrm{lin}$ and the $\sigma^+\sigma^-$ cases. We have not used a Gaussian beam in our simulations because it is too computationally expensive. In addition, the difference between experiment and theory might be reduced by using new calculations of higher order interactions. We are currently elaborating the calculation to enhance the accuracy.

\section{\label{sec:5}Conclusions}

We have presented an experimental study of the modulation transfer spectroscopy of the D\textsubscript{1} transition in potassium. We have also presented a theoretical model which can be used to predict the MTS signal which shows generally good agreement with the experimental measurements. We have shown that for the D\textsubscript{1} transition, the MTS signal shows strong features originating from ground-state crossover transitions. We have optimized various experimental parameters, including the beam intensities and polarizations, to maximise the $X(1,2)\rightarrow F'=2$ feature, providing a good reference for laser frequency stabilisation. We expect these results will be of interest to researchers employing the D\textsubscript{1} transition for laser cooling and optical pumping of potassium in quantum gas experiments. 

\section*{Acknowledgements}

We acknowledge insightful discussions with Sarah Bromley, Jonathan Mortlock and Phil Gregory. This work was supported by the UK Engineering and Physical Sciences Research Council (EPSRC) Grant EP/P01058X/1 and the National Research Foundation of Korea (NRF) grant funded by the Korea government (MSIT) (No. 2020R1A2C1005499).

\bibliography{apssamp}

\end{document}